\documentclass{iau}
\usepackage{graphicx}

\title[Binarity and CSEs of Cepheids] 
{An interferometric view on binarity and circumstellar envelopes of Cepheids}

\author[A. Gallenne et al.]   
{A.~Gallenne$^1$,
P.~Kervella$^2$, A.~M\'erand$^3$, J.D.~Monnier$^4$, J.~Breitfleder$^{2,3}$,
G.~Pietrzynski$^{1,5}$ \and W.~Gieren$^1$}

\affiliation{$^1$Universidad de Concepci\'on, Departamento de Astronom\'ia, \\
Casilla 160-C, Concepci\'on, Chile \\ email: {\tt agallenne@astro-udec.cl} \\[\affilskip]
$^2$LESIA, Observatoire de Paris, CNRS UMR 8109, UPMC,\\ Universit\'e
  Paris Diderot, 5 Place Jules Janssen, F-92195 Meudon, France \\[\affilskip]
$^3$ European Southern Observatory, Alonso de C\'ordova 3107,\\  Casilla 19001, Santiago 19, Chile \\[\affilskip]
$^4$Astronomy Department, University of Michigan,\\ 1034 Dennison Bldg, Ann Arbor, MI 48109-1090, USA  \\[\affilskip]
$^5$Warsaw University Observatory, Al. Ujazdowskie 4, 00-478 Warsaw, Poland
}

\pubyear{2014}
\volume{301}  
\pagerange{1--2}
\jname{Precision Asteroseismology}
\editors{J.A. Guzik, W.J. Chaplin, G. Handler \& A. Pigulski, eds.}
\begin{document}

\maketitle

\begin{abstract}
Optical interferometry is the only technique giving access to milli-arcsecond (mas) resolution at infrared wavelengths. For Cepheids, this is a powerful and unique tool to detect the orbiting companions and the circumstellar envelopes (CSE). CSEs are interesting because they might be used to trace the Cepheid evolution history, and more particularly they could impact the distance scale. Cepheids belonging to binary systems offer an unique opportunity to make progress in resolving the Cepheid mass discrepancy. The combination of spectroscopic and interferometric measurements will allow us to derive the orbital elements, distances, and dynamical masses. Here we focus on recent results using 2- to 6-telescopes beam combiners for the Cepheids X~Sgr, T~Mon and V1334~Cyg.
\end{abstract}

\firstsection 
\section{Introduction}


Cepheid CSEs are interesting for several aspects. Firstly, they might be related to the past or ongoing stellar mass loss, and might be used to trace the Cepheid evolution history. Secondly, their presence may bias distance determination using Baade-Wesselink methods, and then bias the calibration of the IR period--luminosity (P--L) relation. 

Galactic Cepheids in binary systems are also important to measure fundamental stellar parameters. The dynamical masses can be estimated, providing new constraints on evolution and pulsation theory. This would give a new insight on Cepheid masses, and might settle the discrepancy between pulsation and evolution masses. Binary systems are also valuable for obtaining independent distances to Cepheids, needed to calibrate the P--L relation.


\section{Circumstellar envelopes}
The characterization of CSEs is essential as they give access to the present mass-loss rate of Cepheids. The CSEs were probably formed through past or ongoing mass loss, possibly generated by shock waves in the pulsating atmosphere of Cepheid. Our recent work using the mid-IR instrument VLTI/MIDI (2-beams combiner) enabled us to use the radiative transfer code \texttt{DUSTY} to model the CSEs of X~Sgr and T~Mon (\cite[Gallenne et al.~2013b]{Gallenne_2013_09_0}). The fitted models gave mass-loss rates in the range $10^{-7}$--$10^{-8} M_\odot$\,yr$^{-1}$, consistent with the expected theoretical range. We also estimated a relative IR excess in agreement with our previous work (\cite[Gallenne et al.~2011]{Gallenne_2011_11_0}), and derived a mid-IR correlation between the relative excess and the pulsation period of Cepheids. This correlation shows that longer-period Cepheids have larger IR excesses.

In the near-IR the CSE flux emission might be negligible compared to the photospheric continuum, but this is not the case in the mid-IR, where the CSE emission dominates. This has an impact on the calibration of the P--L relation in the thermal domain (e.g. \cite[Monson et al.~2012]{Monson_2012_11_0}), while the bias in the $K$ band still needs to be more deeply studied.

\section{Binary Cepheids}
Most of Cepheid companions are located too close to a Cepheid ($\sim1$--40\,mas) to be spatially resolved with a single-dish telescope. We are engaged in a long-term interferometric observing program, that aims at detecting and characterizing the companions of nearby Cepheids. Our main objectives are the determination of accurate masses and geometric distances from high-precision astrometry. The derived empirical masses will provide very valuable constraints in modeling the pulsation and evolution of intermediate-mass stars. 

\cite[Gallenne et al.~(2013a)]{Gallenne_2013_04_0} presented the first results of this program for the Cepheid V1334~Cyg, using the CHARA/MIRC instrument (6-beams combiner). The companion was clearly detected in the closure phase signal. We combined our astrometric measurements with spectroscopic data to derive the complete set of orbital elements. We were also able to estimate the lower limit for the mass and distance.

Recently, the companion of AW Per and AX~Cir was also detected using the IR combiner CHARA/MIRC and VLTI/PIONIER, respectively, and the results will be published soon.

\section{Conclusion}
Thanks to the high angular resolution provided by interferometry, we are able to study the close environment of Cepheids. It is important to keep analyzing the effect of the CSEs on the near- and mid-IR P--L relation, and probe the origin of their presence. Empirical masses of Cepheids are also particularly important to constrain theoretical models. Binary Cepheids are the best tools for such measurements. Interferometry coupled with radial velocity measurements seems to be a powerful way to estimate Cepheid masses.

\vspace{.5cm}
\textbf{Acknowledgments.} AG acknowledges support from FONDECYT grant 3130361. JDM acknowledges funding from the NSF grants AST-0707927 and AST-0807577. WG and GP gratefully acknowledge financial support for this work from the BASAL Centro de Astrof\'isica y Tecnolog\'ias Afines (CATA) PFB-06/2007. Support from the Polish National Science Centre grant MAESTRO DEC-2012/06/A/ST9/00269 and the Polish Ministry of Science grant Ideas Plus (awarded to GP) is also acknowledged.

\end{document}